# Mechanism of Discordant Alternans in Spatially Homogeneous Tissue.

short title: theory of discordant alternans

Mari A. Watanabe, MD, PhD; Flavio H Fenton*, PhD;
Steven J Evans, MD**; Harold M Hastings, PhD*; Alain Karma, PhD

Physics Department
Dana Building
Northeastern University
Boston, MA 02115
Telephone: 617-373-2929
Fax: 617-373-2943
karma@presto.physics.neu.edu

* Mathematics Department, Hofstra University, Hempstead, NY

** Cardiology Department, Beth Israel Hospital, New York, NY

**abstract**

Discordant alternans, the phenomenon of separate cardiac tissue locations exhibiting action potential duration (APD) alternans of opposite phase, appears to be a potential mechanism for electrocardiographic T wave alternans, but its initiation mechanism is unknown.  We studied behavior of one- and two- dimensional cardiac tissue spatially homogeneous in all respects, including APD restitution and conduction velocity restitution, using the Beeler Reuter ion channel model.  We found that discordant alternans were initiated when spatial gradients of APD arose dynamically, such as from fixed rate pacing of a cable end (sinus node scenario), or from fixed rate pacing at one site preceded by a single excitation wavefront from another site (ectopic focus scenario).  In the sinus node scenario, conduction velocity restitution was necessary to initiate discordant alternans.  Alternating regimes of concordant and discordant alternans arose along the length of the cable, with regimes delimited by nodes of fixed APD.  The number of observable nodes depended upon pacing rate and tissue length.  Differences in beat to beat conduction velocity values at steady state were small.  In the ectopic focus scenario, variable conduction velocity was not required for induction of discordant alternans.  In both scenarios, stability of node position was dependent upon electrotonic coupling.  Other mathematical models produced qualitatively similar results.  We conclude that spatial inhomogeneities of electrical restitution are not required to produce discordant alternans;  rather, discordant alternans can arise dynamically from interaction of APD and conduction velocity restitution with single site pacing, or from APD restitution alone in two site pacing.

**Introduction**

Occurrences of electrocardiographic T wave alternans (TWA) were noted soon after the first electrocardiograms were published 90 years ago, but such overt TWA was rare enough to elicit case reports [1]. Overt TWA was frequently associated with profibrillatory conditions such as angina, acute myocardial ischemia [reviewed in 2] and long QT syndrome [3]. Recent advances in signal processing techniques have led to the ability to detect microscopic (microvolt level) TWA from the body surface [4] and to the discovery that microscopic TWA is a fairly common phenomenon at higher heart rates. Clinical studies show a high correlation between large amplitude microscopic TWA and sudden cardiac arrest [5, 6]. Commercial instruments can now quantify microscopic TWA in cardiology offices, and one such test was cleared by the FDA last year as the first non-invasive test to identify patients at risk for sudden cardiac death [7].

Many experimental studies have been conducted to elucidate the mechanism of electrocardiographic alternans. Some have focused on alternans of action potential duration (APD) and mechanical contraction [reviewed in 8] that occurs at fast heart rates. Others have studied the correlation between alternans and ischemia [reviewed in 2], or long QT syndrome [3, 9, 10]. A recent optical mapping study by Pastore et al strongly suggests that TWA is related to discordant alternans (two spatially discrete sites exhibiting APD alternans of opposite phases) [11], a phenomenon described previously between epicardial and endocardial electrocardiograms [12]. The purpose of our study was to describe the origins of discordant alternans in spatially homogeneous cardiac tissue, using mathematical modeling and non-linear dynamics theory. Previous theoretical work has already suggested that sustained APD alternans [13] arises when the slope of APD restitution (the dependence of APD on duration of electrical diastole, i.e., diastolic interval or DI) is greater than 1.

**Methods**

A discrete model was used to illustrate initiation mechanisms of discordant alternans. It was composed of a 100 cell cable with cell-cell distance of 0.0375 mm, and no electrotonic coupling. Each cell adhered to the APD restitution equation APD [ms] = 300-250*exp(-DI/100) and to the intercellular conduction velocity restitution equation CV [mm/ms] = 0.025-0.02*exp(-DI/40) or = 0.025. To simulate long term behavior of discordant alternans, we used the Beeler Reuter (BR) ion channel model [14] in a cable or a sheet of cardiac cells. We used the forward Euler method to solve the partial differential equations, with step-sizes of 0.25 mm and 0.02 ms. Restitution of APD and conduction velocity (dependence of conduction velocity on DI) are both monotonically increasing curves in the BR model [15]. To confirm validity of results from the BR model, we also analyzed discordant alternans in the Noble ion channel model [16] and in 2- and 3- variable models in which restitution parameters could be varied [15, 17]. Figures 1 and 2 are from the discrete model, all others are from the BR model.

**Results**

*Basic mechanisms of discordant alternans initiation*

Initiation of discordant alternans requires two conditions to be met, an APD restitution slope greater than 1 at the stimulus site, which allows sustained APD alternans at that site, and an initial spatial variation of DI. Initial variation of DI can be produced even in absence of spatial heterogeneity of APD and conduction velocity restitution curves. To illustrate, we will use a standard geometrical technique called cobwebbing (**figure 1**). Drawing a vertical line from the x axis to the restitution curve APD = f(DI) gives an APD value for a given DI. If time between stimuli is S1S1, then new DI = S1S1 - old APD. Therefore, new DI can also be obtained graphically by drawing a horizontal line from the coordinates (old DI, old APD) on the restitution curve until it intersects the auxiliary line DI + APD= S1S1 at (new DI, old APD). This process of drawing a vertical line to the restitution curve and a horizontal line to the auxiliary line, can be repeated indefinitely to produce sequential DI, APD values. The intersection of the restitution curve and auxiliary line is called the fixed point (DI*, APD*), because the DI and APD are fixed in time at that point. The cobwebbing technique shows that if initial DI < DI*, then the subsequent DI and APD are always longer than the current values (a short-long sequence), and conversely, if initial DI > DI*, then the next DI and APD are always shorter (a long-short sequence). I.e., any spatial gradient of DI in which DI values flank DI* must produce discordant alternans.

*Two scenarios of discordant alternans initiation*

The laddergrams in **figure 2** show two scenarios for producing initial DI gradients. In the ectopic focus scenario (**figure 2A**), we assumed that the tissue was excited uniformly before the first S1 stimulus ($S1_1$), and that depolarization wavefronts traveled with fixed velocity, making time from depolarization to depolarization 290 ms everywhere along the cable. At the stimulus site, the first DI was 62 ms (white space). APD (grey space) was then 165 ms from f(62)=165. The second DI was 125 from 290-165=125. The second APD was then f(125)=228. The third DI was 290-228=62, identical to the initial DI, proving the sequence would repeat itself stably. Farther from the stimulus site, the fact that the wavefront could not travel instantaneously to other sites gave rise to longer DI with distance traversed. Increasing DI produced increasing APD. This in turn led to a reversal in spatial distribution of DI when the wavefront from $S1_2$ travelled down the cable. Therefore, at the stimulus site, APD alternated with a short-long-short-long sequence, while at the bottom end of the cable, the same stimuli produced a long-short-long-short sequence, or discordant alternans. Cell 19, where DI equaled DI*, separated the concordant and discordant alternans regimes. Below the bottom edge of the laddergram, conduction block occurred when APD grew to equal S1S1.

With sufficient time, APD values came to alternate between 165 and 228 ms for cells 0-18, and cells 20-100 with different phases, while cell 19 remained at APD*. This unphysiological end state of a sharp 63 ms APD difference between two cable segments was due to absence of electrotonic coupling in this discrete model.

In the sinus node scenario, similar to the stimulation protocol of ref [11], stimuli were introduced following long quiescence (**figure 2B**).  There was only one pacemaker site, but conduction velocity was allowed to vary with DI.  Short DI (2 ms) produced by $S1_2$ at cell 1 produced slow conduction.  Slow conduction gave rise to longer DI as the wavefront traveled down the cable, which in turn sped up conduction and shortened DI, thereby producing sinusoidal DI gradients.  The conduction velocity restitution itself produced the initial variation of DI, which spanned the DI* of 91 ms.  APD node position where the APD was identical for two consecutive beats gradually moved towards the stimulus end.  For example, the first node formed just beyond cell 100 and node APD value was approximately 167 ms. The node was next seen at cell 49, with node APD value of 207.  Subsequent nodes were seen at cells 35 and 28 respectively.  The remainder of the results section describes the dynamics of discordant alternans in this sinus node initiation scenario, using the BR model.

*Spatial gradients of APD, cycle length and conduction velocity*

APD gradients evolve in time.  How the gradients grow in magnitude until the even and odd beat gradients intersect is shown in **figure 3** for a cable length of 80mm, basic cycle length (BCL) of 310 ms.  The intersection (node) asymptotically approaches a location near the pacing site.  The final state is shown in **figure 5A**.

**Figure 4** shows the steady state distribution of voltage over the length of a 1-D cable for a full alternans cycle, at a BCL of 310 ms, for two cable lengths.  The vertical axis is time, so rotating these figures 90 degrees clockwise produces the laddergram orientation.  The voltage tracings in the left panel exhibit several features.  The two depolarization wavefronts look roughly linear, similar to the laddergram, indicating that fluctuation of conduction velocity is relatively small. APD and DI can be visualized from high voltage and low voltage segments along the vertical axis separated by densely apposed voltage traces. The upstroke of the tenth voltage trace from the bottom shows the approximate location where APD is fixed in time. To the left of this position, APD alternates concordantly with the stimulus end, and to the right, discordantly. It is important to note that some of the voltage distributions exhibit a minimun and maximum. The significance of this phenomenon is discussed later in relation to electrotonic interactions. The **right panel** shows voltage distribution when 3 nodes exist at steady state.

**Figures 5 A-C** show even and odd steady state APD gradients for 3 different tissue lengths at a BCL of 310 ms. There are 1, 2 and 3 nodes for cables of length 80, 117.5 and 135 mm, respectively. Multiple nodes imply alternating regimes of concordant and discordant alternans. If a node exists for a particular tissue length, it occurs at the same position regardless of cable length. However, the cable must be considerably longer than where a node is expected, for the node to exist at all. E.g., there is only one node for the 80 mm cable, although judging from the 117.5 mm cable, two nodes might be expected. This is because the node first arises where the spatial APD gradients cross, moves in the direction of the stimulus site, and becomes fixed in position. Even though a node can exist or "fit" in a particular length of tissue, the tissue has to be long enough for the node to be born. Where the node is born depends on the conduction velocity restitution curve. E.g., increased conduction velocity moves the initial APD node position farther from the stimulus site, and slope must not be zero for the gradients to cross.

**Figures 5 C-E** show the relationship between steady state APD, cycle length, and conduction velocity gradients for a cable length of 135 mm. At the APD nodes, cycle length alternates by approximately 6 ms. At the cycle length nodes where cycle

length is fixed, APD alternates, replicating the condition at the stimulated end of the cable. Conduction velocity nodes are closely but not exactly aligned with the APD nodes, and alternate by approximately 50 mm/s.

*Relationship between BCL and node position and number*

In experiments [11], shortening BCL causes concordant tissue to become discordant. The division of length, BCL parameter space into regions of 0 (concordant) through 4 node alternans in **figure 6A** illustrates a mechanism for this phenomenon. Above a BCL of 320 ms, there is no permanent alternans of APD at the stimulus site. Below a BCL of 280 ms, stimulus rate is high and produces refractory block of every other stimulus. BCL between these two threshold values produces stable alternans of APD at the stimulus site, and concordant or multi-nodal discordant alternans in the tissue as a whole. As tissue length increases, more nodes can exist for a given BCL. For a fixed tissue length, the number of nodes can increase by one as BCL is shortened. E.g., at a tissue of length 30 mm, alternans is concordant at or above a BCL of 290 ms. Shortening BCL to 285 ms or lower produces one node, i.e., one region of discordant alternans, similar to experiments. Electrocardiograms computed from an electrode placed above the center of a 30 mm cable for BCL of 290 and 285 ms are shown at the bottom of the figure. Both electrocardiograms show some alternans of QRS. The lower electrocardiogram exhibits T wave alternans.

**Figure 6B** shows steady state APD node position versus BCL for a cable length of 140 mm. Nodes move closer to the paced end of the cable and internodal distances shrink as cycle length is reduced.

*Restitution curve during alternans*

The outer graph in **Figure 7** shows the relationship between the standard restitution curve of the BR model (dotted line) and the two DI, APD relationship curves that arise dynamically during discordant alternans. Both of the dynamic curves are lower than the standard restitution curve for most DI, and higher for a small range of short DI. A particular DI value can produce two APD values, as shown by the curling ends of the dynamic restitution curves. The split of the dynamic curves shows why DI can alternate, yet still produce the same APD. The inset similarly shows standard and dynamic conduction velocity restitution curves. It is important to note that the different dynamic restitution curves do not signify an *a priori* spatially heterogenous distribution of restitution curves, but a dynamic modulation of inherent restitution properties by electrotonic coupling.

*2-D simulation and results from other models*

**Figure 8** shows temporal evolution of APD gradient for alternate beats in a 2-dimensional sheet in the sinus node scenario. The APD node line, where APD was identical on consecutive beats, moved towards and stabilized near the pacing site. APD gradient on the diagonal line of the 2-D sheet is shown at the bottom.

Simulations of discordant alternans in other models [15-17] produced results qualitatively similar to simulations from the BR model, demonstrating the robustness of mechanisms of initiation and evolution across different models. One quantitative difference was noted. In the BR model, QRS alternans and T wave alternans arose simultaneously, due to similar DI over which APD and conduction velocity restitution had large slopes, while in the other models, T wave alternans could precede QRS alternans with rate increase.

**Discussion**

The first condition that must be satisfied to initiate discordant alternans is to have non-transient alternans at the pacing site. In theory, a restitution curve with a maximum slope greater than 1 will produce stable alternation of APD for some range of BCL [13,18,19], although there are experiments showing some discrepancy from theory [20], and some experimental [21,22] and theoretical [23] results ascribe a more important role to slope of dynamic rather than standard restitution curve in alternans production. The APD restitution curve of the BR model has a maximum slope that is greater than 1, and stable alternans was produced at the pacing site in accordance with theory. Steep (>1) restitution curve slope has also been shown to produce spiral wave break-up [15,24] in mathematical models, a phenomenon believed to be an analog for ventricular fibrillation. The slope of restitution curve is thus an indirect link between ventricular fibrillation and alternans, and may be related to the close relationship between TWA and ventricular fibrillation.

The second condition necessary to produce discordant alternans was a non-uniform initial distribution of DI in space. We found two ways to accomplish this in electrically homogeneous tissue. In the ectopic focus scenario, conduction velocity could be fixed, but required two stimulus sites (**figure 2A**). That situation is analogous in a 2 dimensional system to a straight edge depolarization wavefront coming from one direction, followed by straight edge depolarization wavefronts coming from a second direction, such as would happen when a sinus beat is followed by beats from a regularly firing ectopic focus, or the case where a ventricular premature contraction is followed by sinus beats. In the sinus node scenario, conduction velocity restitution produced spatial gradients of DI with only one stimulus site.

The boundary between discordant and concordant alternans was represented by nodes where APD was identical on two consecutive beats. In the sinus node scenario, nodes formed, gradually moved towards the stimulus site, and stopped at a distance determined by BCL and cable length. Multiple nodes indicated multiple regimes in space where the phase of APD alternans was alternately concordant or discordant with the phase at the stimulated end. The number of nodes seen depended on tissue length. Although only single nodes have been observed experimentally [11], the simulation predicts that more may be found if larger tissue sizes are studied. Simulations replicated the experimental observation of faster pacing converting concordant to discordant alternans.

During discordant alternans, there was alternans not only of APD, but of DI, cycle length, and conduction velocity. The relative importance of APD alternans and conduction velocity alternans has been debated, with some giving alternans of conduction velocity the primary role [25] and others, alternans of APD [26-28]. Two recent studies support primacy of APD alternans; activation times were identical from beat to beat [29, 30]. The results of our simulations show that both APD and conduction velocity alternans are important. It is possible that the magnitude of conduction velocity alternans is too small (50 mm/s compared to an overall velocity of 450 mm/s) to produce measurable differences of activation time in experiments.

The difference of the dynamic restitution curves from the standard restitution curve was found to be due to electrotonic coupling of cardiac tissue. This is best explained by studying **Figure 4**. When DI is very short, a minimum appears in the spatial distribution of voltage with the next depolarization. APD at the voltage minimum is longer than expected from the restitution curve, because electrotonic current flows towards the minimum, keeping voltage elevated. Conversely, when DI is long, a maximum appears in the spatial distribution of voltage with the next depolarization. APD at the voltage maximum is shorter than expected from the restitution curve, because electrotonic current flows away from the maximum.

The split between the two dynamic curves themselves was also found to be due to electrotonic coupling. There existed an asymmetry of electrotonic coupling with respect to propagation direction for a given APD gradient. For example, consider three adjacent segments in a cable, *a*, *b*, and *c*, with DI-dependent APD values $APD_a$, $APD_b$, and $APD_c$, in the uncoupled state. If the excitation sequence is *a -> b -> c*, segment *a* supplies electrotonic current during its phases 1-3 to segment *b*, while segment *c* supplies electrotonic current during its phases 0-3 to segment *b*. If the excitation sequence were reversed, segment *b* would receive phase 1-3 current from segment *c* and phase 0-3 current from segment *a*. The sum of electrotonic currents passing from segments *a* and *c* to segment *b* during $APD_b$ is therefore different depending on depolarization direction, even if the uncoupled scalar values $APD_a$ and $APD_c$ remain the same. This asymmetry can be quantified rigorously, and is part of a separate manuscript in progress. The split of dynamic restitution curves due to electrotonic coupling is directly responsible for stability of APD node position, and therefore of discordant alternans.

Based on our results, we expect that spatial heterogeneity in the form of smooth gradients of electrical restitution would not change initiation and evolution characteristics of discordant alternans. However, effects of discontinuous tissue heterogeneities such as non-conducting scar tissue require further study.

In summary, discordant alternans arises from the simple cardiac tissue characteristics of APD restitution and conduction velocity restitution. It does not require spatially heterogeneous tissue properties. Previous theory predicts that concordant alternans could be prevented by reducing heart rate, or alternatively by reducing APD restitution curve slope, such as by calcium channel antagonists [21,30 and references therein]. Results from the present study suggest that discordant alternans, and by inference, TWA, might be prevented by pharmacological agents that reduce conduction velocity restitution slope, even in presence of alternans.

**Acknowledgments**

This study was supported by American Heart Association Grant in Aid 96009660 and NIH SCOR in Sudden Cardiac Death 1PL50HL-52319.

**Figure legends**

**Figure 1** Cobweb diagram showing mechanism of discordant alternans initiation. The curve represents APD restitution equation APD = 300-250*exp(-DI/100), the -45° line represents equation DI+APD=290. The vertical line shows DI*, the fixed point value of DI. The L-shaped arrows illustrate how small value DI and APD to the left of DI* are followed by large value DI and APD to the right of DI*, and vice versa. The square represents the stable alternation between DI, APD pairs (62, 165) and (125, 228).

**Figure 2.** Laddergrams showing two scenarios of discordant alternans initiation. The vertical axis represents 100 cells of a cable, and the horizontal axis represents time. Stimuli were applied at the top end of the cable. Bands of white and grey depict electrical diastole and action potential respectively. Boxed numerical values indicate DI and APD. Bold lines depict depolarization wavefronts. Computation of DI and APD was independent of values at adjacent points, i.e., lacked electrotonic effects. **A**: Ectopic focus scenario. The entire cable had depolarized and repolarized before $S1_1$. Conduction velocity was fixed, but increasing DI with distance from stimulus end provided the DI gradient necessary for discordant alternans initiation. **B**: Sinus node scenario. The cable was in quiescence before $S1_1$. Conduction velocity restitution provided the DI gradient necessary for discordant alternans initiation. The cell number at which APD values (in boxes) were the same for two consecutive beats gradually decreased during initiation of discordant alternans.

**Figure 3** Evolution of APD gradients in sinus node scenario, cable length 80 mm, BCL 310 ms. The gradient for even (hashed) and odd (solid line) beats are shown for stimulus numbers 2,3 (**A**), 30, 31 (**B**), 50, 51 (**C**) and 150, 151 (**figure 5A**). The APD node (gradient intersection) moved leftwards with increasing stimulus number.

**Figure 4** Steady state distribution of voltage over a cable for a full alternans cycle at a BCL of 310 ms, for cable lengths of 80 (**left**) and 135 mm (**right**). Voltage traces are plotted every 6 ms with upward shift for clarity. Arrows indicate time of stimulation.

**Figure 5**  Steady state APD, cycle length, and conduction velocity gradients over a cable for even (hashed) and odd (solid line) beats. BCL was 310 ms. **A, B, C**: APD node position and number of APD nodes at 3 cable lengths 80, 117.5 and 135 mm. **C, D, E**: The relationship between APD gradient, cycle length gradient, and conduction velocity gradients for a cable length of 135 mm.

**Figure 6**  Relationship between APD node position, number of nodes, BCL, and cable length at steady state. **A**. The parameter space is divided into zero (concordant), 1, 2, 3 and 4 node regimes. As cycle length is shortened for a given cable length, the number of nodes can increase. Electrocardiograms computed for BCL of 290 and 285 ms at tissue length of 30 mm are also shown. **B**. Node position as a function of cycle length for cable length of 140 mm. As cycle length was reduced, internodal distances decreased.

**Figure 7**  The relationship between the standard (dotted line) and dynamic restitution curves during discordant alternans, (hashed line - even, solid line - odd beats), in a cable 62.5 mm long, for BCL of 305 ms. Large graph: APD restitution curves. Inset: Magnified view of conduction velocity restitution curves.

**Figure 8**  APD gradients in an 80 mm x 80 mm 2-dimensional sheet produced by a point stimulus applied at the bottom left corner every 310 ms. **Left to right - Top row**: APD gradient for stimulus number 2, 20 and 52. **Center row**: APD gradient for stimulus numbers 3, 21 and 53. **Bottom row**: APD gradients on the diagonal of the 2-D sheets. The black lines in the 2-d sheets indicate APD node line position where APD was fixed for consecutive beats. When stimulation was continued, the APD node line became a quarter circle (a line equidistant from stimulus site).

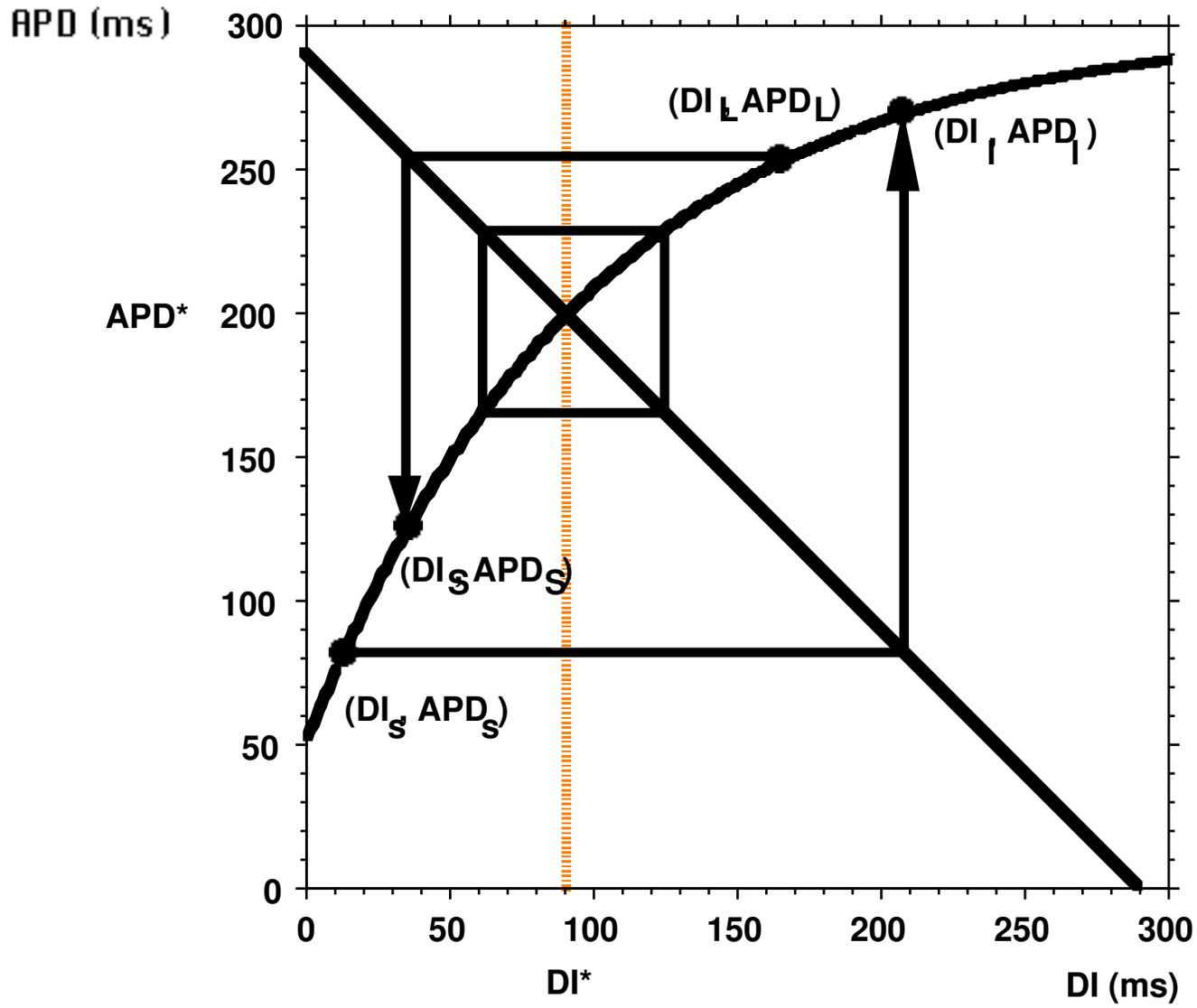

Figure 1

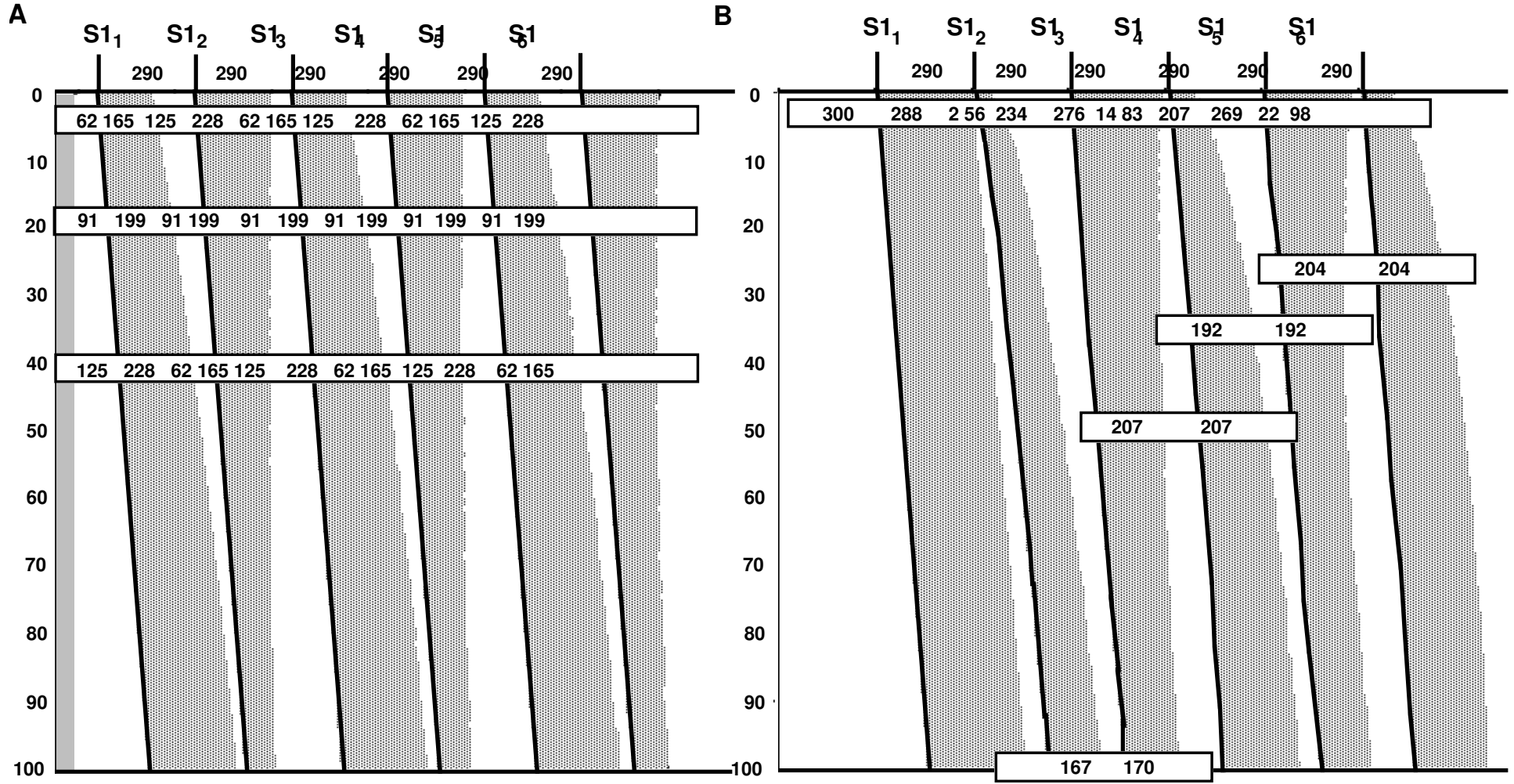

Figure 2

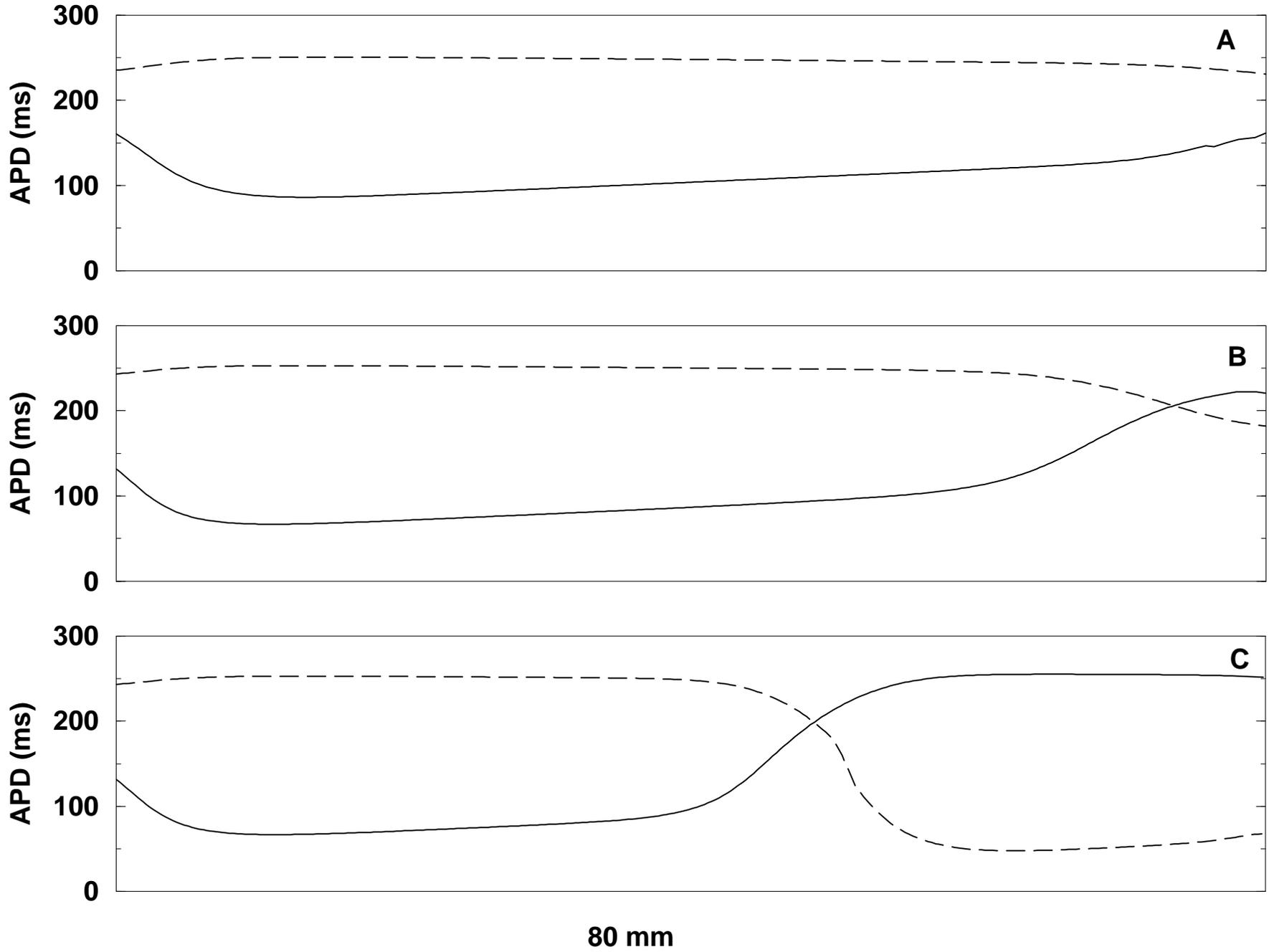

Figure 3

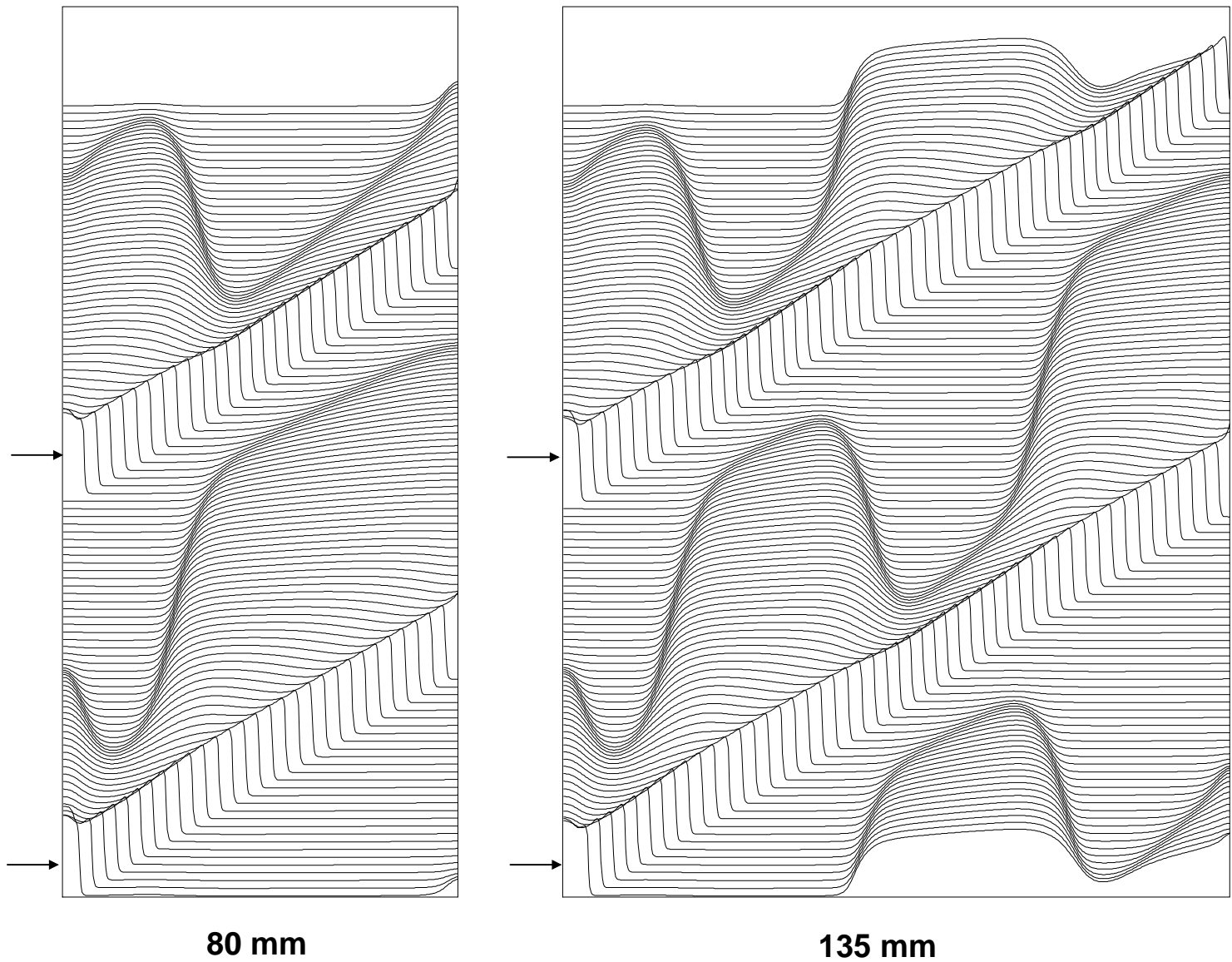

**80 mm**  **135 mm**

Figure 4

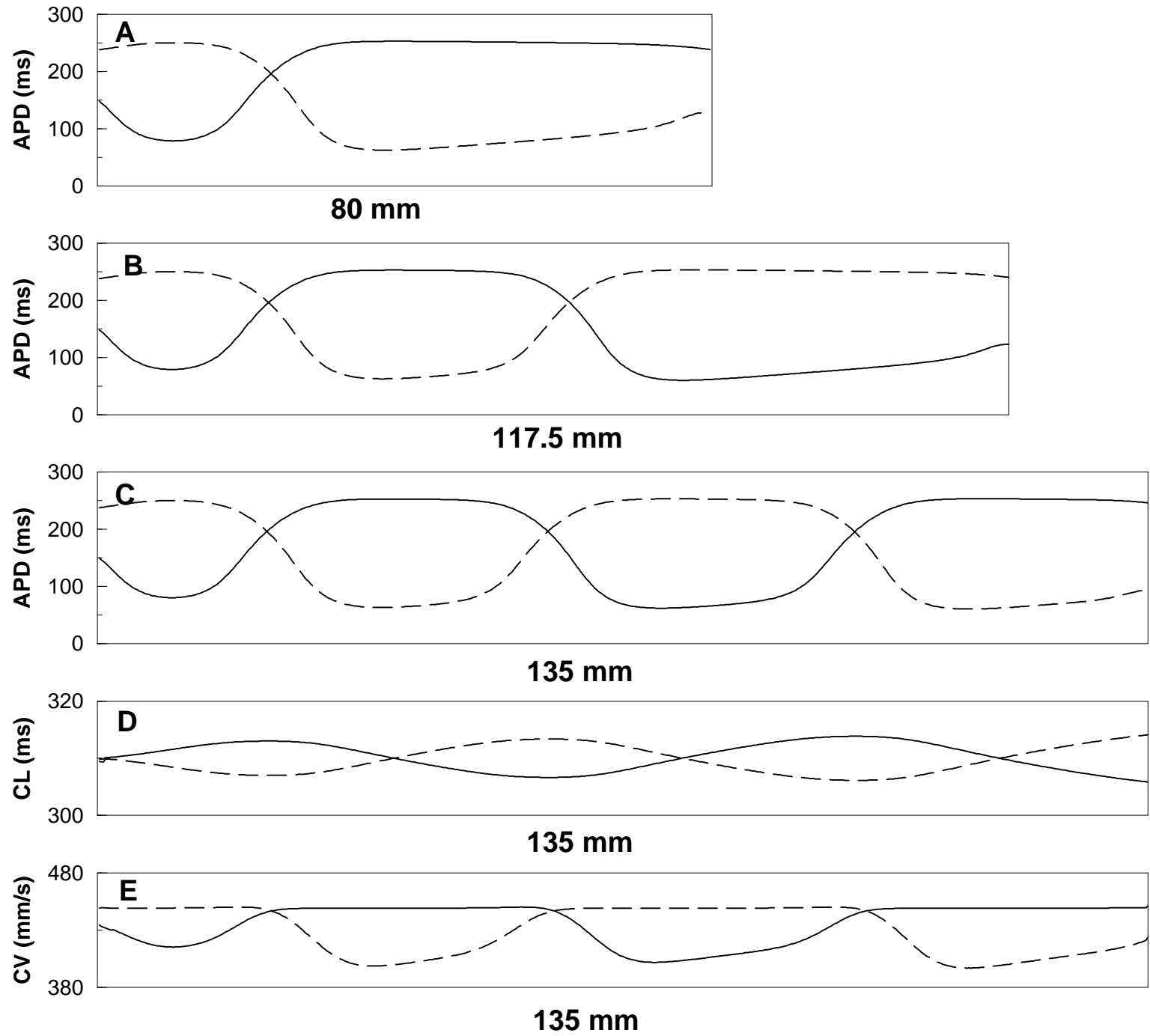

Figure 5

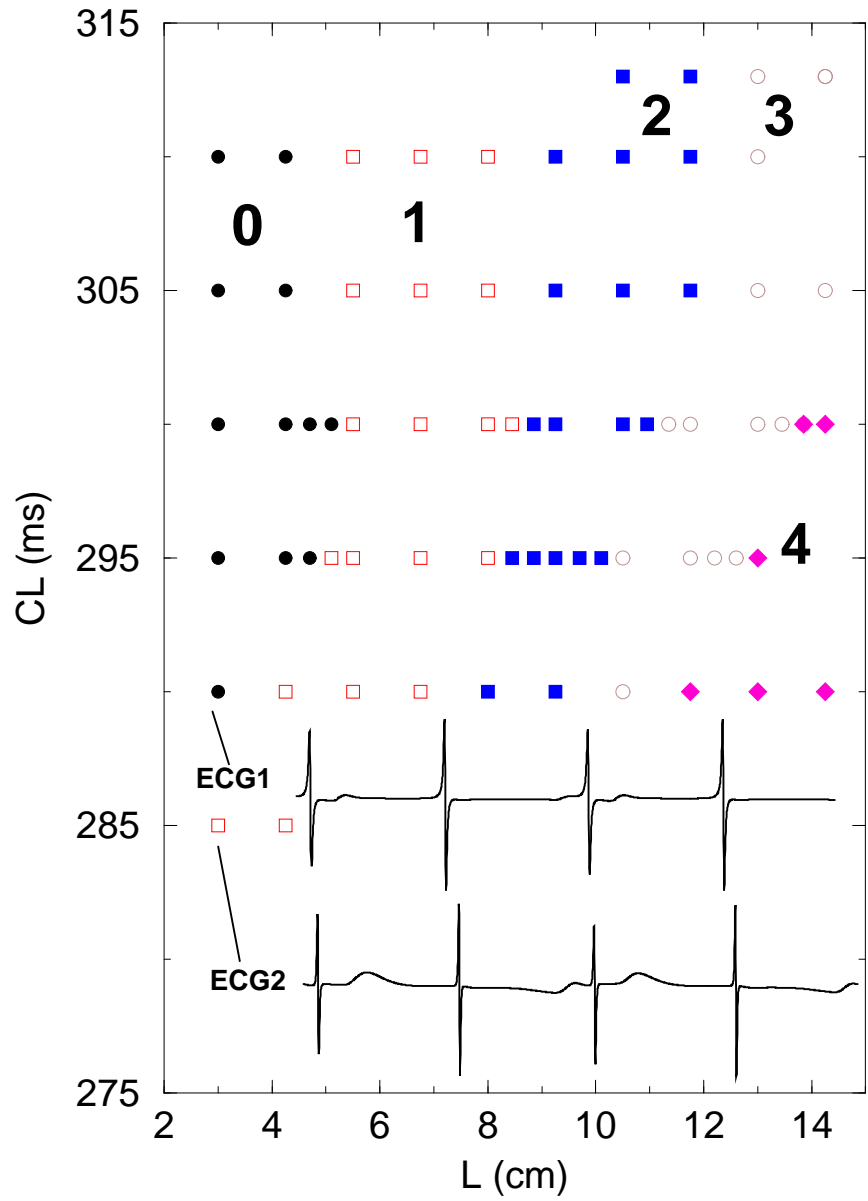
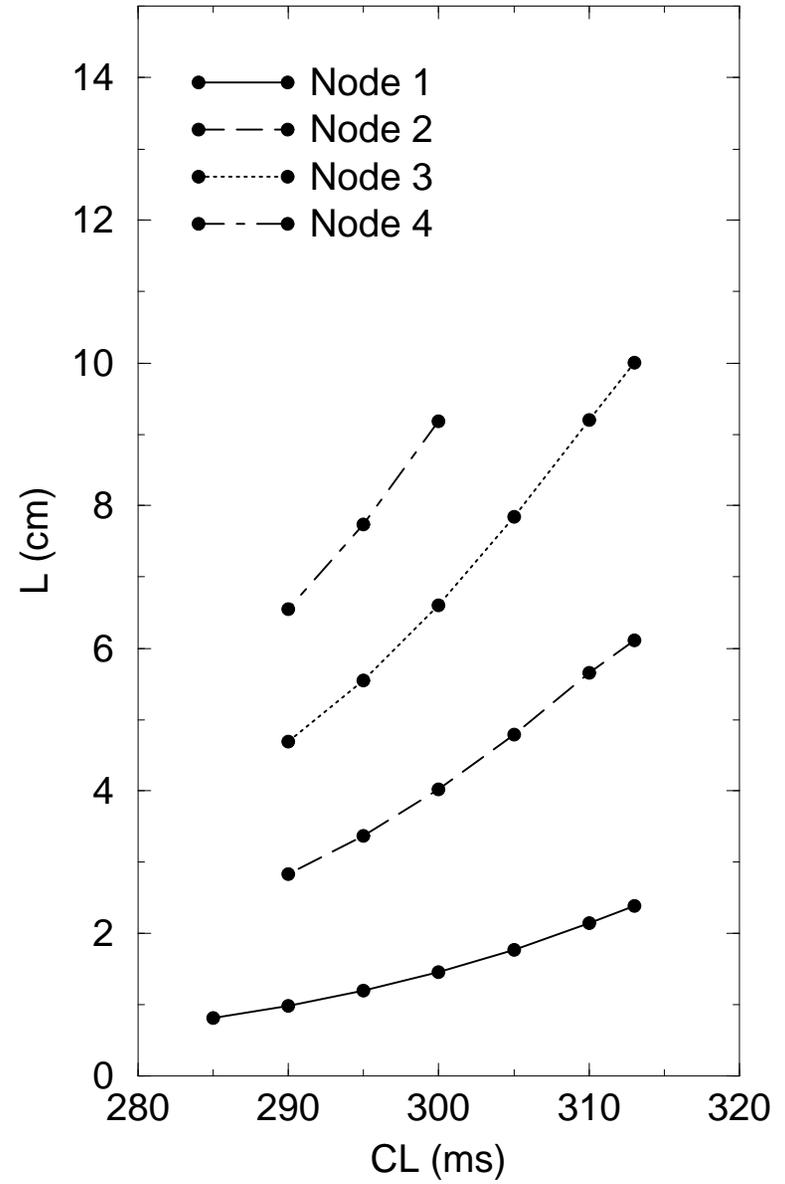

Figure 6

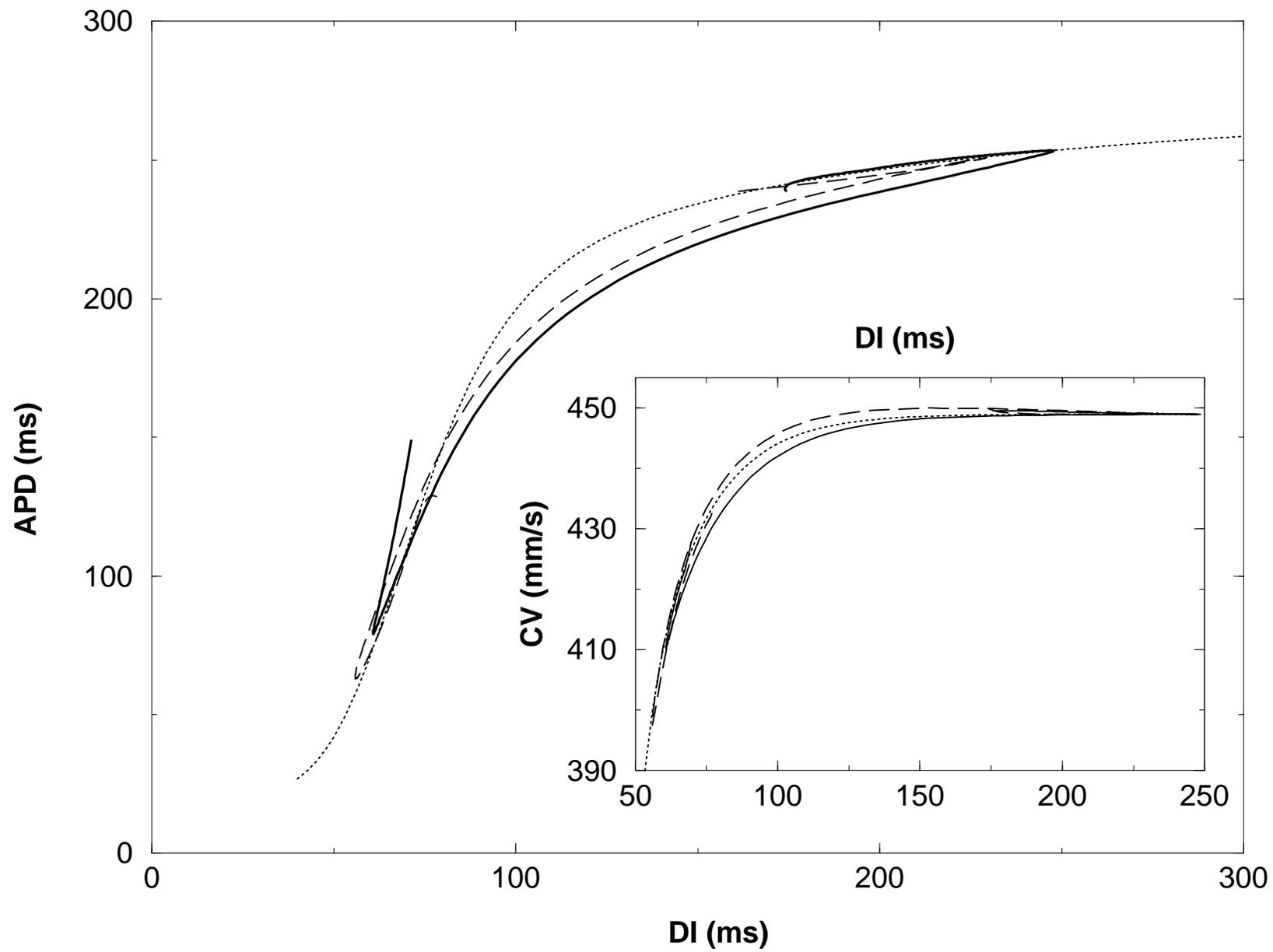

Figure 7

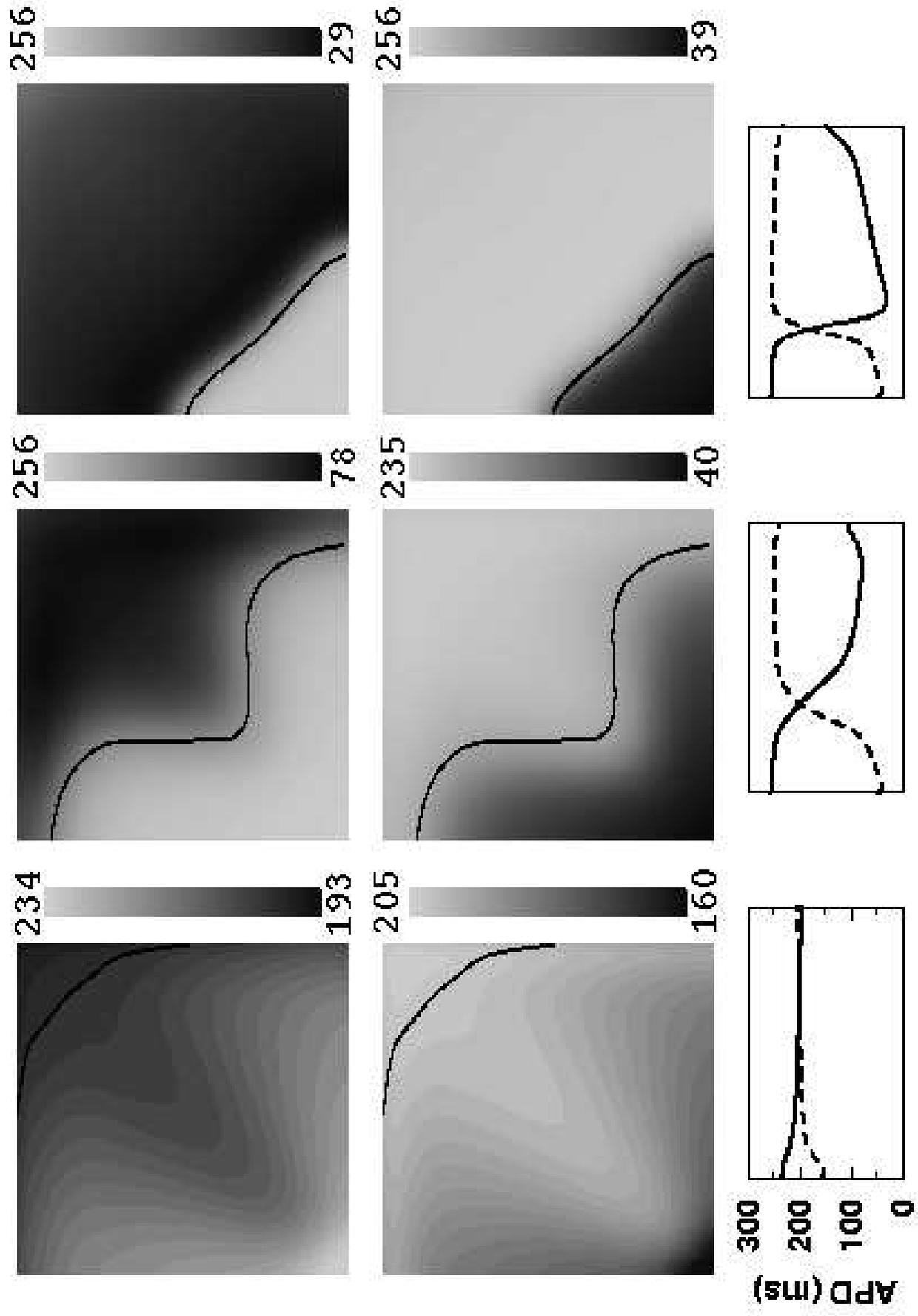

Figure 8

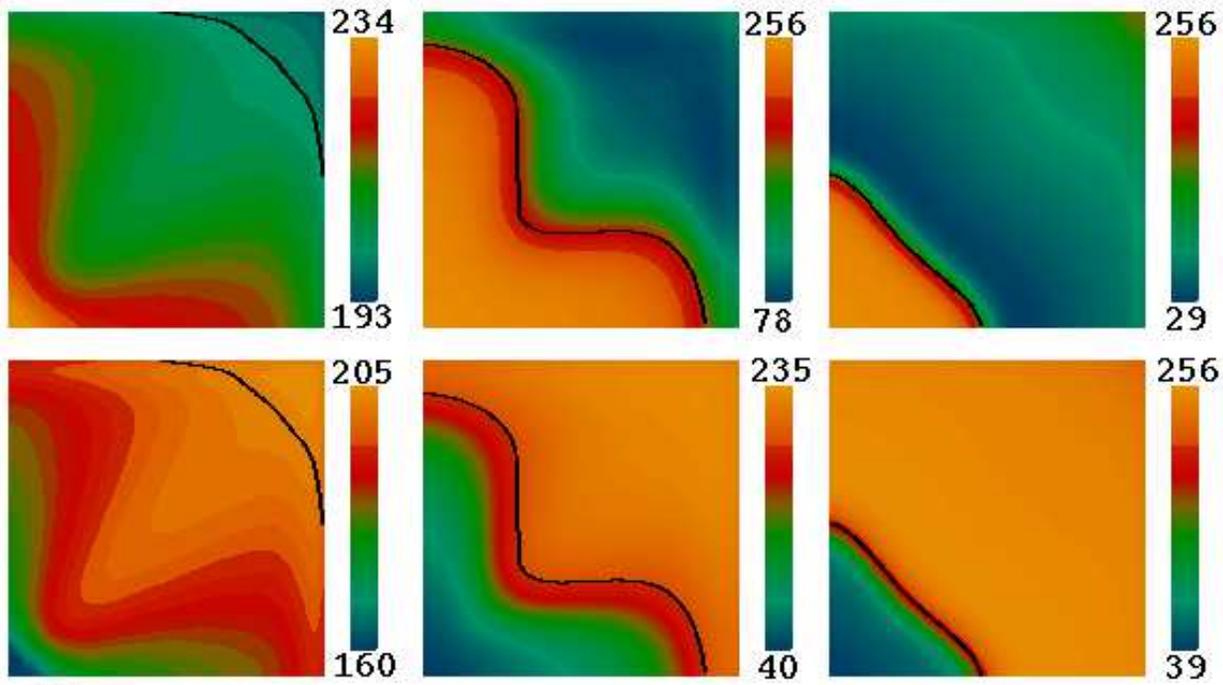